\documentstyle[12pt]{article}

\newcommand{\rf}[1]{(\ref{#1})}

\renewcommand{\thefootnote}{\fnsymbol{footnote}}

\newcommand{\newsection}{    
\setcounter{equation}{0}
\section}
\def\appendix#1{
  \addtocounter{section}{1}
  \setcounter{equation}{0}
  \renewcommand{\thesection}{\Alph{section}}
  \section*{Appendix \thesection\protect\indent \parbox[t]{11.715cm} {#1} }
  \addcontentsline{toc}{section}{Appendix \thesection\ \ \ #1}
  }

\def\bI{{\rm I}}

\def \four{{\textstyle {1\ov 4}}}
 
\def\bi {\bibitem}
\def \ep{\epsilon}

\def \om {\omega}

\def \F {{\cal F}}

\def \del {\partial}

\def \ha{{\textstyle{1\over 2}}}

\def \a {\alpha}
\def \b {\beta}
\def \chi {\chi}
\def \s {\sigma}

\def \m {\mu}
\def \n {\nu}

\def \td {\tilde }
\def \d {\delta}
\def \ci {\cite}

\def \inv {^{-1}}
\def \ov {\over }
\def \four{{\textstyle{1\over 4}}}
\def \fourth{{{1\over 4}}}

\def \e { e }

\def \V {{\cal V}}

\def \pa {\Vert}
\def \hal{{{1\over 2}}}

 \def \G {{\Gamma}}
 
 \def \L {{ \Lambda}}
 \def \d {\delta}
\def \foot{\footnote}

\def \Tr {{\rm Tr }}
\def\np {{  Nucl. Phys. }}
\def \pl {{  Phys. Lett. }}

\def \prl {{  Phys. Rev. Lett. }}
\def \pr  {{ Phys. Rev. }}

\def \bi{\bibitem}

\def \tr {{\rm tr }}
\def \W  {{\cal  W}}
\def \VV {{\rm V}}

\def\det{\hbox{det}}
\def\be{\begin{equation}}
\def\ee{\end{equation}}
\def\beq{\begin{equation}}
\def\eeq{\end{equation}}
\def\bea{\begin{eqnarray}}
\def\eea{\end{eqnarray}}
\def\LB{\left (}
\def\RB{\right )}
\def\lsb{\left [}
\def\rsb{\right ]}
\def\te{\theta}
\def\de{\partial}

\def\xo{{\bar X}^{(1)}}
\def\xt{{\bar X}^{(2)}}
\def\xijo{{\bar X}^{(1)}_{ij}}
\def\xijt{{\bar X}^{(2)}_{ij}}
\def \la{\label}
\def \ci{\cite}

\def \FI{\td \F}
\def \f{\td f}
\def \ff {{\rm  f}}
\def \N {{\cal N}}
\def \NN {{\bf N}}
\def \FF {{\rm F}}
\def \i{{i}}

\begin{document}
\begin{titlepage}
\begin{flushright}
ITEP--TH--19/97\\
Imperial/TP/96-97/47\\
hep-th/9705120\\
\end{flushright}
\vspace{.5cm}

\begin{center}
{\LARGE   Interactions of type IIB D-branes  \\[.3cm]
 from  D-instanton matrix model  }\\[.2cm]
\vspace{1.1cm}
{\large I. Chepelev${}^{{\rm 1,}}$\footnote{E-mail: guest@vxitep.itep.ru}
and A.A. Tseytlin${}^{{\rm 2,}}$\footnote{Also at Lebedev Physics
Institute, Moscow. \ E-mail: tseytlin@ic.ac.uk} }\\
\vspace{18pt}
${}^{{\rm 1\ }}${\it Institute of Theoretical and Experimental Physics,
Moscow 117259, Russia}\\
${}^{{\rm 2\ }}${\it Blackett Laboratory,   Imperial College, London SW7 2BZ, U.K.}
\end{center}
\vskip 0.6 cm

\begin{abstract}
We compute long-distance interaction potentials between certain
1/2 and 1/4 supersymmetric D-brane configurations of type IIB theory,
demonstrating detailed agreement between classical supergravity
and one-loop instanton matrix model results. This confirms the
interpretation of D-branes as described by classical matrix model
backgrounds as being `populated' by large number of D-instantons,
i.e. as corresponding to non-marginal bound states of branes of
lower dimensions. In the process, we establish precise relation
between matrix model expressions and non-abelian  $F^4$
terms in the super Yang-Mills effective action.
\end{abstract}

\end{titlepage}
\setcounter{page}{1}
\renewcommand{\thefootnote}{\arabic{footnote}}
\setcounter{footnote}{0}

\newsection{Introduction}

The aim of this paper
 is to discuss   interactions  between some
D-branes in  type IIB matrix  model of \ci{ikkt}
(see also \ci{peri}).
Our approach  will be  that of  \ci{dbi} where the
$D=10$ $U(N)$  super Yang-Mills theory reduced to a point
was not related to the  (Schild form of) type IIB string action
 as in  \ci{ikkt,li}
but was  interpreted   as the  direct D-instanton counterpart of  the
D0-brane matrix model of \ci{bfss}.
The
two matrix models  can be put  into  correspondence
using T-duality in the time direction.

The Dp-brane configurations in the instanton matrix
 model can be described
\ci{ikkt,dbi,CMZ,fas,FMOSZ}
in  a similar way as in  the 0-brane  matrix model \ci{bfss,grt,bss}.
As was pointed out in \ci{dbi}, they should be identified
not with `pure'  type   IIB  D-branes but
with  D-branes `populated' by  large number
 of D-instantons
 just like
D-branes  in the matrix model of   \ci{bfss}
are `populated' by large number of 0-branes  \ci{lifmat, lif3}.

In what follows we shall confirm  this interpretation
by demonstrating that the corresponding  long-distance
interaction potentials computed
in the  matrix model   and in supergravity
are in precise agreement.
The matrix model (SYM)  result  is the same as the short-distance
limit of the 1-loop open string theory amplitude
while the supergravity  result is the long-distance limit
of the tree-level closed string theory potential.
They agree in the  $N\to \infty$ limit
in which the brane configurations  become supersymmetric
 for the same reason as in the 0-brane matrix model \ci{bfss,lifmat}.

The $U(N)$ SYM theory reduced to a point describes
a collection of $N$ D-instantons \ci{wit2,ggu}.
When  some of  the ten  euclidean dimensions are compactified on a torus $T^{p+1}$,
the  classical backgrounds represented by  constant
abelian  fluxes ($[A_m,A_n] = iF_{mn}$)
correspond  \ci{dbi}
 to 1/2 supersymmetric non-marginal bound states
of type IIB  Dp-branes (i.e. $1+\i$, $3+1+\i$, $5+3+1+\i$)
wrapped over the dual torus $\td T^{p+1}$.
The configuration with self-dual strength  $[A_m,A_n]$ represents
the 1/4 supersymmetric marginal bound state of D3-brane and D-instantons
which we shall denote as  $ 3\pa\i\ $  \ci{ggut,tsee}.

There is  a close  T-duality
relation to similar configurations in 0-brane matrix model
 \ci{grt,bss,lifmat, lif3}.
Indeed, the  interaction
potentials between such D-branes
in the instanton matrix model computed  below
 are direct counterparts of the  corresponding
results in  M(atrix) theory  found in \ci{bfss,ab,lifmat,lif3}
for interactions between
 1/2 supersymmetric branes and in
  \ci{CT} for interactions involving  1/4 supersymmetric  branes.

We shall consider two examples:

 (i) interaction
between   D-instantons and
1/2 supersymmetric  `Dp-branes', i.e.
non-marginal  $ p+(p-2)+\cdots +1+\i$ bound states;

(ii)
interaction
between   `D-string', i.e. $1+\i$ bound state,
and
1/4 supersymmetric
marginal 3-brane--instanton bound state $3\pa \i$.

In section 2 we shall determine the
corresponding closed string theory (supergravity) potentials
using classical probe method (see \ci{CT} and refs. there).
In section 4 we shall reproduce the same expressions
by a  one-loop calculation in the
 instanton matrix model.
In section 3 we shall present  some  general  results
about 1-loop effective action in $D \leq 10$ SYM theories
and explain  their relation  to   the
matrix model computations of  the
 leading  terms in long-distance interaction
potentials.

One  natural   generalisation
of the  present work  is to 1/8 supersymmetric bound states
probed by  D-instantons or other type IIB `D-branes'.
In particular, one may  consider D-brane configurations
corresponding to $D=5$  black holes
as in  \ci{dps,mall} and  \ci{lim,dvv,hal}.

\newsection{ Closed string theory (supergravity)  description}
\subsection{D-instanton --
`Dp-brane'
  interaction
}
To determine the D-instanton--`Dp-brane'
 interaction potential
we shall consider the latter, i.e. the
 $p+(p-2)+\cdots +1+\i$ bound state
of
type IIB D-branes  ($p=-1,1,3,5$)
as a probe moving  in the
classical  D-instanton background.\foot{By
the potential we shall mean the interaction part
of the euclidean action.
The euclidean
time coordinate may be  assumed
to belong to the internal $p+1$-dimensional  torus.
Alternatively,  one  may consider the $p$-branes discussed below
as being  `$(p+1)$-instantons' \ci{ggut},
with  the time coordinate being  orthogonal
to the internal torus.}
This  probe  can be described, as in \ci{CT},
by  the standard
  Dp-brane
  action  with  a constant  world-volume gauge field  background.
The relevant terms in the euclidean  Dp-brane action are
($m,n=1,...,p+1; \ i,j=p+2,...,10$)
 \be
 I_p = - T_p\bigg[
\int d^{p+1} x    e^{-\phi} \sqrt {\det\big( G_{mn } +
 G_{ij} \del_m X^i \del_n X^j +  \F_{mn} \big)}
 - \int_{p+1}  \sum_k C_{2k} \wedge e^{\cal F} \bigg]\ ,
    \la{act}
    \ee
    where $\F_{mn} \equiv  T\inv  F_{mn}$  (in what follows $B_{mn}=0$)
   and
    $C_{2k}$ is the RR $2k$-form potential.
We used the static gauge
    and took  the target-space metric in the  block-diagonal form.
 In general, Dp-brane tension is \ci{pol}
   \be
  T_p \equiv n_p \bar T_p=  n_p g\inv (2\pi)^{(1-p)/2}  T^{(p+1)/2} \ , \ \ \ \ \ \
   T\equiv (2\pi \a')^{-1} \ .
  \la{tens}
  \ee
 We shall  assume that the  euclidean world-volume
 of a type IIB
  Dp-brane is  wrapped over
 a (rectangular) torus  $T^{p+1}$
  with volume $V_{p+1}=(2\pi)^{p+1}R_1...R_{p+1}$
  and that there is a constant
 world-volume gauge field background
 \begin{equation}
 \F_{mn}=\left(
 \begin{array}{ccccc}
 0 & f_1 & & & \\
 -f_1 &0& & & \\
  & &\ddots & & \\
 &&&0 & f_{l} \\
 &&&-f_{l}&0 \\
 \end{array}
 \right) , \ \ \ \ \ \   l\equiv \ha (p+1)\  .
 \label{fmunu}
 \end{equation}
The Dp-brane with the  flux~\rf{fmunu} on its
world-volume  represents
the  non-marginal bound state $(p+(p-2)+\cdots +1+\i)$ of
D-branes of dimensions $p,p-2,...$  \ci{ggut}\
 (with branes of `intermediate' dimensions
being wrapped over different cycles of the torus).  The  total numbers of
branes  of each type  are
\be
n_{p-2} =  n_p  2\pi T \sum_{k=1}^l  f_k R_{2k-1}R_{2k}\,,~~ ...~~\,,\ \ \ \
n_{-1} =   n_p  V_{p+1} \prod_{k=1}^l(  {Tf_k\ov 2\pi }) \,,
\label{number}
\ee
as  can be read off from the Chern-Simons terms
in the D-brane action \rf{act} \ci{doug}.

The D-instanton background
   `smeared' in the directions of the torus $T^{p+1}$
  ($x_1=y_1,\
...,\ x_{p+1}=y_{p+1}$)  is  \ci{ggp}\foot{We   use  the symbol `$\i$' and
subscript `$-1$' to denote  D-instantons and the
corresponding quantities.}
   \be
   ds^2_{10} = H_{-1}^{1/2}(dy_1^2 +\cdots +dy_{p+1}^2+  dx_idx_i ) \ ,
    \la{two}    \ee
$$ e^{\phi} = H_{-1} \ , \ \ \ \ \  C_0 = H_{-1}^{-1}-1  \ , \ \ \ \
H_{-1} = 1 + {Q^{(p+1)}_{-1}\over r^{7-p}} \ , \ \ \ \ r^2=x_i x_i\  .    $$
We shall use the notation $Q^{(n)}_p$  for the coefficient in the harmonic
function $H_p = 1 + {Q^{(n)}_p\over r^{7-p-n}}$
of p-brane  background which is smeared in $n$ transverse toroidal directions.
In general,
\be
Q_p =  N_p g (2\pi)^{(5-p)/2} T^{(p-7)/2} (\om_{6-p})\inv  \ , \ \ \  \ \ \ \
 \om_{k-1} = 2 \pi^{k/2}/\Gamma({ k/ 2})  \ ,
\la{qq}
\ee
\be
Q^{(n)}_p = N_p g (2\pi)^{(5-p)/2} T^{(p-7)/2} (V_n \om_{6-p-n})\inv
  = N_p N_{p+n}\inv  Q_{p+n} (2\pi)^{n/2} T^{n/2} V_n\inv ,
\la{sme}
\ee
where  $V_n$ is the  volume of the flat internal  torus.

  Substituting  the  background  \rf{two}  into the  Dp-brane action
\rf{act}  and  ignoring the dependence of
$X_i$ on world-volume coordinates $x_m$ (so that
  the matrix under the  square root in \rf{act}  becomes
$H_{-1}^{1/2} \d_{mn} + \F_{mn}$)  we find
 \be
 I_p  = -T_p V_{p+1}  \bigg[ H_{-1}^{-1}
\prod_{k=1}^l
 \sqrt { H_{-1} + f_k^2  }- (H_{-1}^{-1} -1) \prod_{k=1}^l  f_k  \bigg]  \,. \la{acti}
\ee
Defining the  `interaction  potential'  $\V (r)$\  ($ r^2= X_i X_i$)
as the deviation from the
 `free' action
of the non-marginal $p + ...+\i$ bound state,
\be
I_p=I_p^{(0)} - \V = -T_pV_{p+1}\prod_{k=1}^l \sqrt{1+f_k^2}-\V    \ ,
\la{vot}
\ee
we get  for the leading long-distance  term in $\V $
\be
\V = \frac{1}{r^{7-p}}Q_{-1}^{(p+1)}T_pV_{p+1} \ \prod^{{l}}_{m=1}
\sqrt{1+f_m^2} \
\bigg[ \sum^l_{k=1}  \frac{1}{2(1+f_k^2)} +\prod^l_{k=1} \frac{f_k}{\sqrt{1+f_k^{2}}}
-1\bigg] +O( \frac{1}{r^{2(7-p)}}) \ .
\la{saa}
\ee
The coefficient here is
\be
Q_{-1}^{(p+1)}T_pV_{p+1}
= 2^{3-l}\ (3-l)! \ T^{l-4} \ n_p N_{-1}\ \ , \ \ \ \ \ \  p=2l-1 \ .
\ee
In the limit of the  large  background field $\F_{mn}$ \ ($f_k\gg 1$),
i.e. for  large instanton `occupation number' $n_{-1}$  \rf{number},
 we   find  (we assume that $l \le 3$ and set $T=1$)\foot{Let
 us note that the  subleading $\frac{1}{r^{2(7-p)}}$  term  in $\V$ \rf{vot}
is proportional to (cf. \rf{saa})
$$
\prod^l_{m=1} \sqrt{1+f_m^{2}} \ \bigg[
  \sum^l_{k=1}  \frac{1}{2(1+f_k^2)}
 +\prod^l_{k=1} \frac{f_k}{\sqrt{1+f_k^{2}}}
- { 1\ov 8} \big(\sum^l_{k=1}  \frac{1}{1+f_k^2}\big)^2
+ \fourth  \sum^l_{k=1}  \frac{1}{(1+f_k^2)^2 }\bigg]  . $$
The leading term in  the large field ($f_k \to \infty$) expansion
of  this expression   vanishes.
}
\be
\V = -\frac{1}{r^{8-2l}}  2^{-l} (3-l)!\  n_p N_{-1}\
\prod^{{l}}_{m=1} f_m \
\lsb 2\sum^{{l}}_{k=1}  f_k^{-4} -\big( \sum^{{l}}_{k=1} f_k^{-2}\big)^2\rsb + ... \, .
\la{resu}
\ee
For example, in the case of $p=1$, i.e. the  D-instanton--`D-string'
interaction
\be
\V= - \frac{1 }{r^{6} }  n_1 N_{-1} \f^3  + ... \ , \ \ \ \   \ \
 \f\equiv  f^{-1}_1 \ .
\la{exa}
\ee
Note that the potential \rf{resu}
vanishes for $p=3$ and $f_1=f_2$.
In this case the background field $\F_{mn}$ is self-dual
and the interaction between  D-instanton
and $3+1+i$ non-marginal bound state becomes essentially  the same
as the  interaction between D-instanton
and $3\pa i$ marginal bound state\foot{D-instanton does not
couple to D-string charge; the contribution
 of the latter is in any case suppressed for large $f_k$.}
 but `$i - (3\pa i)$' is
 a BPS configuration \ci{ggut}. Analogous  conclusion  is  reached
in the T-dual case of  0-brane -- $4+2+0$ bound state interaction: when
the  magnetic flux on 4-brane is self-dual,
$0 - (4+2+0)$ interaction is the same as the $0 - (4\pa 0)$ one
 \ci{CT}.

The expression \rf{resu} can be put  in the following `covariant' form
\be
\V = -\frac{1}{r^{8-2l}}  2^{-l} (3-l)!\  n_p N_{-1}\
\sqrt{\det\ \F_{mn} }  \
\bigg[   \FI_{mk}\FI_{kn}\FI_{ns}\FI_{sm} - \fourth
(\FI_{mn} \FI_{mn})^2 \bigg] + ... \, ,
\la{rsu}
\ee
 $$
 \FI_{mn} \equiv (\F_{nm})^{-1} \ .  $$
Since  the D-instanton number  in \rf{number}
is equal to
\be
n_{-1} = n_p (2\pi)^{-l} V_{2l} \sqrt{\det\ \F_{mn} } \ ,
\la{nuum}
\ee
we can  represent \rf{rsu} also as
\be
\V = -\frac{ (3-l)!\ \td V_{2l}     }{(4\pi)^l r^{8-2l}} \     {n_{-1}} N_{-1} \
\bigg[   \FI_{mk}\FI_{kn}\FI_{ns}\FI_{sm} - \fourth
(\FI_{mn} \FI_{mn})^2 \bigg]
 + ...  \ ,
\la{rsut}
\ee
where $\td V_{2l}$ is the volume of the dual torus,
\be
   V_{2l} \td V_{2l} = ({2\pi \ov T})^{2l} =  ({2\pi})^{2l} \ .
\la{voo}
\ee
The  $\FI^4$ coefficient
in this  expression   is  exactly the same  as the  quartic
 term in the
expansion of the Born-Infeld action  $\sqrt{\det (\d_{mn} + \FI_{mn})}$
or in the open string effective action.
This can be seen directly from  \rf{acti}
by noting that the expression there is 
$ H_{-1}^{-1}[  \sqrt { \det ( G_{mn} + \F_{mn}) } - \sqrt {  \det  \F_{mn}} ]=
\sqrt {  \det  \F_{mn}} \bigg( H_{-1}^{-1} [ \sqrt { \det ( \delta_{mn} +  H^{1/2}_{-1} \F^{-1}_{mn}) } - 1] \bigg) $.
The reason for this non-trivial  coincidence
(note that $\FI_{mn}$ is the {\it inverse} of
the background field $\F_{nm}$ in the probe action)
 will become clear  below
when we reproduce \rf{rsut} from the matrix model.

\subsection{Interaction of  `D-string'   with
  3-brane--instanton bound state
}

To determine the interaction potential
between the  non-marginal bound state of D-string  and D-instanton
and the marginal bound state of D3-brane and D-instanton
we shall consider  $1 +\i$ as a probe moving in the $3\pa \i$ background.
As above,
the  action for the $1+\i$  probe will be
 the D-string action \rf{act}
with a  constant flux \rf{fmunu} on 2-torus
representing the D-instanton charge.

 The $3\pa i$  type  IIB  supergravity background \ci{tsee}
is T-dual to $4\pa 0$ or $5\pa 1$ solutions \ci{ts2}.
We shall assume that the 3-brane world volume
is wrapped around 4-torus (in directions $1,2,3,4$)
and that the world volume of $(1+\i)$-brane
    probe is parallel to $(5,6$) directions,
i.e. that the world volumes do not share
common dimensions.\foot{Here the adequate  interpretation  is that
the time direction is
orthogonal to  both of the $(1+i)$
and  $(3\pa i)$ world-volumes
  \ci{ggut}.}
The corresponding metric, dilaton and RR scalar fields
 smeared in  the  $(5,6)$ directions are  \ci{ts2}
   \be
   ds^2_{10} = (H_{-1}H_3)^{1/2}  [
  H^{-1}_3 (dy_1^2 + ...+ dy_4^2)  +  dy_5^2 + dy_6^2 +
    dx_i dx_i ] \ ,
    \la{mee}
    \ee
$$ e^{\phi} = H_{-1}  \ ,  \ \  \ \ \  \ \
C_0 = H_{-1}^{-1}-1  \ ,
\ \ \ \ \ \   H_{-1}= 1 + {Q^{(6)}_{-1}\over r^2} \ , \ \ \ \ \ \
H_3= 1 + {Q^{(2)}_3\over r^2} \ ,   $$
where $Q^{(n)}_p$ are given by \rf{sme}\
($C_2=0$; the value of $C_4$  background
will not be important below).
Ignoring the dependence on derivatives of $X_i$  we find for
 the  `D-string'   probe action $I_{1}$
 ($f\equiv f_1$)
$$
I_{1}  = -T_1  \int d^2 x
  \bigg[ H_{-1}^{-1}  \sqrt { H_{-1}H_3  + f^2   }
    - (H_{-1}^{-1} -1) f   \bigg]
$$
\be
    = -T_1 V_2 f    \bigg[ 1 + H_{-1}^{-1} \bigg(
\sqrt { 1  + H_{-1}H_3
f^{-2}  } - 1\bigg)   \bigg]  \equiv   -T_1 V_2 \sqrt { 1 + f^2} - \V
 \ .  \la{accq}
\ee
The leading   long-distance  interaction term in  $\V$
 is
\be
\V =   \frac{1}{2r^2} T_1 V_2 \sqrt { 1 + f^2}
\bigg[ Q^{(2)}_3  { 1 \ov 1 + f^2 } -
Q^{(6)}_{-1} \bigg( 1 -  { f \ov \sqrt { 1 + f^2}} \bigg)^2  \bigg] +  O({1 \ov r^4})
\ .
\la{iio}
\ee
This  expression is in direct T-duality correspondence
with the static
potential  between the $2+0$ and $4\pa 0$ bound states
in \ci{CT}.

The potential \rf{iio}
has the following large $f$ (large instanton charge
$n_{-1}$  of $1+i$) expansion,
cf.\rf{exa}
\be
\V = \frac{1}{2r^2} n_1  \bigg(N_3 \f -  N_{-1}\pi^2 V^{-1}_4\f^3
-\ha N_3 \f^3\bigg) + ...
  \ , \ \ \ \ \   \f\equiv f\inv \ ,
\la{eex}
\ee
where $ V_4$ is the volume of the  4-torus.\foot{The large $f$ limit of the $1\ov r^4$
subleading term in $\V$ is
$ - { 1\ov 8 r^4} T_1 V_2 Q^{(2)}_3 ( 2 Q^{(6)}_{-1} +  Q^{(2)}_3 ) f^{-3}$.}
$\V$ can be expressed
in terms of
\be n_{-1} =n_1 (2\pi)\inv  V_2 f = n_1 2\pi \td V_2\inv \td f\inv    \  \la{nuu}
\ee
as follows
\be
\V = \frac{\td V_2    }{4\pi r^2}\   {n_{-1}}\  \bigg(N_3 \f^2 -
  {\textstyle { 1\ov (4\pi)^2} }  N_{-1}   \td V_4\f^4
-\ha N_3 \f^4\bigg) + ...
   \ .
\la{erx}
\ee
Since in the matrix model representation
$N_3$ will be the
 instanton number of a  gauge field on the dual 4-torus,
\rf{erx}  will be, like \rf{rsut},  proportional
 to the integral of  $F^4$ terms over the dual 6-torus
($ N_3 \f^4$ will be a subleading  correction).

\newsection{One-loop effective action in $D\leq 10$ super Yang-Mills theory}
To put  matrix model computations in a proper perspective,
it  is useful
 to give  a summary of some general  results about
  the  one-loop  effective action $\G(A)$ of
 maximally supersymmetric YM theory in $D\leq 10$ dimensions.
\subsection{UV divergences and `large mass' expansion}
In general,
\be
\G = \ha \sum_{a} c_a \ln \det \Delta_a = -\ha \int^\infty_{\L^{-2}} {ds \ov s}
\ \tr \sum_a c_a e^{-s \Delta_a} \ ,
\ee
where the sum over $a$ runs over bosonic, background gauge ghost and fermionic
contributions taken with appropriate relative coefficients
 ($c_a=1,-2,-\fourth$). \
$\Delta_a$ are second order differential operators ($ - D^2 + {\cal X} $)
depending on background value of the gauge field and $\L\to \infty$ is
UV cutoff.
The divergent part of $\G$  can be expressed in terms of the DeWitt-Seeley
coefficients ${\ \bf b}_n$
\be
 (\tr\ e^{-s \Delta})_{s\to 0}\simeq {1\ov (4\pi)^{D/2}}  \sum^{\infty}_{n=0}
s^{n-D \ov 2} \int d^D x\ {\bf b}_n (\Delta)
\ ,
\la{use}
\ee
i.e.
\be
\G^{(\infty)} = - {1\ov (4\pi)^{D/2}} \int d^D x \bigg(
{\L^D \ov D} {\ \bf b}_0 + {\L^{D-2}  \ov D-2} {\ \bf b}_2 + {\L^{D-4}  \ov D-4} {\ \bf b}_4
+ ... + \hal \ln \L^2 {\ \bf b}_D \bigg)  \ ,
\ee
where ${\bf b}_n\equiv \sum_{a} c_a {\bf b}_n (\Delta_a)$.
For pure YM  theory \ci{fraa} \ ${\ \bf b}_4 = {1\ov 12} (D-26) \Tr F^2_{mn}$ (the appearance of the
coefficient $D-26$  can be understood from string theory
 \ci{mets}), while  for  $D=10$ SYM theory and its reductions to
lower dimensions \ci{fraa}
\be
{\ \bf b}_0={\ \bf b}_2={\ \bf b}_4={\ \bf b}_6=0 \ ,
\ee
so that SYM theories in $D\leq 7$  are one-loop   UV finite.
At the same time, ${\ \bf b}_8$ and ${\ \bf b}_{10}$ are, in general, non-vanishing.
In particular,  in a constant abelian background
${\bf b}_{10}=0$ but  ${\ \bf b}_8\sim F^4 \not=0$,
implying the presence of  logarithmic divergence in $D=8$ SYM and quadratic
divergence in $D=10$ theory  \ci{fraa}.
The general non-abelian expressions for ${\ \bf b}_8$  and ${\ \bf b}_{10}$
 (up to $F^5$ terms) in SYM theory
 were found in \ci{mets} (basing on the results of \ci{vdv})
\be
{\ \bf b}_8 = {{\textstyle{2\over 3}}} \Tr \bigg(F_{mk} F_{nk} F_{mr} F_{nr}
+ \ha  F_{mk} F_{nk} F_{nr}  F_{mr} - \four
F_{mk} F_{mk} F_{nr} F_{nr}  - {{\textstyle{1\over 8}}}
F_{mk} F_{nr}  F_{mk} F_{nr} \bigg) \ ,
\la{bbb}
\ee
$$
{\ \bf b}_{10} =   -  {{\textstyle{1\over 15}}}
   \Tr \bigg(D_q F_{mk} F_{nk} D_q F_{mr} F_{nr}
+ \ha  D_q F_{mk} F_{nk} D_q F_{nr}  F_{mr}$$
\be  - \four
D_q F_{mk} F_{mk} D_q F_{nr} F_{nr}  - {{\textstyle{1\over 8}}}
D_q F_{mk} F_{nr} D_q  F_{mk} F_{nr} \bigg) + O(F^5) \ .
\la{bib}
\ee
The trace  $\Tr$ is in the adjoint
representation\foot{For generators of $SU(N)$
$\Tr(T_aT_b) = N \d_{ab}, \ \ \tr(T_aT_b) = \ha  \d_{ab}$
 and
$\Tr X^2 = 2N \tr X^2 , \ \
\Tr X^4 = 2N \tr X^4 + 6 (\tr X^2)^2$, \ $ X= X^aT_a$\
(see  \ci{okubo};  similar  expressions in
Appendix B of \ci{mets} should be multiplied by factor of 2).
The same relations are true for  a matrix  $X$ belonging to
$U(N)$ algebra  provided  $X$  in the r.h.s. is replaced by
its traceless part $X \to \bar X= X - { 1 \ov N} \tr X\ \bI$.
}
 and we dropped
 gauge-dependent $O(D_m F_{mk})$
terms  which vanish on the  equations of motion.

The reason why the structure of ${\ \bf b}_8$
(i.e. of the coefficient of quadratic divergence in $D=10$ SYM)
is the same as of the $F^4$ term
in the  open superstring effective action
was explained in \ci{mets}.\foot{
This term
  can be extracted from the $\a'\to 0$ limit
of the string
one-loop effective action  (${1\ov \a' } F^4\to   \L^2 F^4$)
    if one includes planar as well as
 non-planar $ (\tr F^2)^2)$ contributions.
The tree-level open string
effective action contains  similar $F^4$ term (the kinematic
factor in the tree-level and 1-loop  4-vector amplitude is
the same \ci{grs}) but
 with  $\tr$ instead of  $\Tr$ \ci{tgw}.}

Let us now  formally shift $\Delta_a$ by the same constant term $M^2$
and define `IR regularised' effective action $\G_M$
\be
\G_M \equiv  \ha \sum_{a} c_a \ln \det ( \Delta_a  + M^2)=
 -\ha \int^\infty_{\L^{-2}} {ds \ov s}  e^{-s M^2}
\tr \sum_a c_a e^{-s \Delta_a}  \ .
\la{gam}
\ee
This modified  1-loop effective action  is   finite
in $D \le 7$ and has the following
 {\it large} $M$ expansion  (we use \rf{use})
\be
\G_M \simeq
 - \ha  \int d^D x \sum_{n=0}^\infty {\G({n-D \ov 2}) \ov (4\pi)^{D/2} M^{n-D} } {\ \bf b}_n
 =    - {1\ov 2 (4\pi)^{D/2}} \int d^D x  \bigg[
  { \G({8-D \ov 2}) \ov  M^{8-D} } {\ \bf b}_8
+    { \G({10-D \ov 2})   \ov M^{10-D} } {\ \bf b}_{10}
 +  ...  \bigg] .
   \la{poop}
\ee
The explicit form of the leading term is
$$
\G_M =  - { (3-\ha D)! \ov 3 (4\pi)^{D/2} M^{8-D} }  \int d^D x
\  \Tr \bigg(F_{mk} F_{nk} F_{mr} F_{nr}
+ \ha  F_{mk} F_{nk} F_{nr}  F_{mr}
$$
\be
- \four
F_{mk} F_{mk} F_{nr} F_{nr}  - {{\textstyle{1\over 8}}}
F_{mk} F_{nr}  F_{mk} F_{nr} \bigg)  + O( {1 \ov M^{10-D}} ) \  ,
\la{tet}
\ee
or, equivalently, in terms of the trace  in the fundamental representation of
$U(N)$
$$
\G_M =
 - { 2 (3-\ha D)! \ov 3 (4\pi)^{D/2} M^{8-D} }  \int d^D x \bigg(\  N
\  {\rm tr}\bigg[\bar F_{mk} \bar F_{nk} \bar  F_{mr} \bar  F_{nr}
+ \ha   \bar F_{mk} \bar  F_{nk} \bar  F_{nr}  \bar  F_{mr}
$$
$$
- \four
 \bar F_{mk} \bar  F_{mk} \bar  F_{nr}  \bar F_{nr}  - {{\textstyle{1\over 8}}}
 \bar F_{mk}  \bar F_{nr}  \bar  F_{mk} \bar  F_{nr} \bigg]  $$ $$
+\   3 \bigg[
{\tr}( \bar F_{mk} \bar  F_{nk})\ \tr( \bar  F_{mr} \bar  F_{nr})
+ \ha  \tr( \bar F_{mk} \bar  F_{nr})\ \tr(  \bar F_{nk}  \bar  F_{mr})
$$
\be
-\  \four \tr( \bar F_{mk}  \bar F_{nr})\ \tr( \bar F_{mk}  \bar F_{nr})
  - {{\textstyle{1\over 8}}} \tr( \bar F_{mk}  \bar F_{mk})\ \tr( \bar  F_{nr}  \bar F_{nr})
 \bigg]
\bigg)  + O( {1 \ov M^{10-D}} )  \ ,
\la{fund}
\ee
where $\bar F_{mn} \equiv F_{mn} - {1 \over N} \tr F_{mn}  \bI $.
This expansion  is useful in discussions of
long-distance interactions between  Dp-branes where  $D= p+1$ and
$M$ is  proportional  to separation  $b$ between branes, i.e.
$M^2 = T b^2$
(expressions  related to special cases of \rf{tet},\rf{fund}  appeared
 in \ci{ikkt,CMZ,ber} and, in particular, in \ci{mall}; see also below).

Note that the subleading $O( {1 \ov M^{10-D}})$
correction determined by   ${\bf b}_{10}$
  vanishes in the case of constant abelian backgrounds
which  describe, e.g.,
 interactions  between
 1/2 supersymmetric non-marginal bound states
of D-branes.
The coefficient  ${\bf b}_{10}$ is, in general, non-vanishing
for non-abelian background fields.


\subsection{Constant  abelian gauge field background}
The  one-loop effective action of SYM theory in $D$
dimensions  can be computed   explicitly
for a constant  abelian gauge field background
(i.e.  for $F_{mn}=\FF^I_{mn} T_I  $ belonging to the Cartan subalgebra of a
compact semisimple Lie algebra) following  \ci{fraa,avr}.
The basis $T_I$ $(I=1, ..., r)$  in the   Cartan subalgebra
in
 the adjoint representation
 can be chosen
as a set of $d\times d$  {diagonal}
  matrices
 $T_I= {\rm diag} (0, ...,0, \a^{(1)}_I, - \a^{(1)}_I, ..., \a^{(q)}_I, - \a^{(q)}_I)$,
where
$\{\a^{(i)}_I\}$
are positive roots ($i= 1, ..., q, \  q=\ha (d-r)$,
\ \ $\sum_{i=1}^q \a^{(i)}_I \a^{(i)}_{I'}= \d_{II'}$).
Let us define $\FF^{(i)}_{mn} = \FF^I_{mn} \a^{(i)}_I$
and assume that all  $\FF^{(i)}_{mn}$ have `block-diagonal' form
(we choose space-time dimension to be  even $D=2l$)\foot{The expressions that follow
are true also in more general case if
the parameters
$\ff^{(i)}_k$  are simply replaced by  Lorentz invariants constructed out of
 $\FF^{(i)}_{mn}$  (separately for each $i$)
according to the rules \ci{avr}:
$\sum^l_{k=1}( \ff^{(i)}_k)^{2h} =  \ha (-1)^h
 \FF^{(i)}_{m_1 n_1} \FF^{(i)}_{n_1 m_2}... \FF^{(i)}_{n_{h-1} m_1}$, \ \  $h=1, ...,l$. }
\begin{equation}
 \FF^{(i)}_{mn}=\left(
 \begin{array}{ccccc}
 0 & \ff^{(i)}_1 & & & \\
 -\ff^{(i)}_1 &0& & & \\
  & &\ddots & & \\
 &&&0 & \ff^{(i)}_{D/2} \\
 &&&-\ff^{(i)}_{D/2}&0 \\
 \end{array}
 \right) ,
 \end{equation}
Then
 one finds the   following  general
   expression
for $\G_M$ in \rf{gam}
($\VV_{D} \equiv \int d^D x$)
$$
 \Gamma_M =- {2\VV_D\ov (4\pi)^{D/2} }
\int\limits_{0}^{\infty }\frac{ds}{s^{1+D/2} }\,\e^{-M^2s}\
  \sum_{i=1}^{q} \bigg(   \prod^{D/2}_{k=1}\frac{\ff^{(i)}_k s }{\sinh \ff^{(i)}_ks}\
$$
\be \times
 \bigg[\sum_{k=1}^{D/2} \LB\cosh 2\ff^{(i)}_ks -1\RB
 -4(\prod^{D/2}_{k=1}\cosh
 \ff^{(i)}_ks-1) \bigg] \bigg) \  .
\la{eeixs}
\ee
In what follows we shall consider the special case  when the background is such that
$\NN$ of $\FF^{(i)}_{mn}$ are equal to the same $\FF_{mn}$
while the rest vanish, i.e. when
 $\ff^{(i)}_k = \ff_k$, \ $i= 1, ..., \NN$.
 The corresponding
 background field strength is  given by diagonal matrices
in the adjoint or fundamental representaions:
$$F_{mn}^{(adj)} =   {\rm diag} (0, ...,0,\FF_{mn} , - \FF_{mn}, ..., \FF_{mn}, - \FF_{mn}) \ , \ \ \ \ \
F_{mn}^{(fund)} =\LB
\begin{array}{cc}
\FF_{mn} \ {\bI}   & 0 \\
 0 &  0  \\
 \end{array} \RB\   , $$
where $\bI$ a unit ${n \times n}$ matrix
and  $\NN= n (N-n)$.
Then
\begin{equation}
 \Gamma_M  =- {2 \NN \ \VV_D\ov (4\pi)^{D/2} }
\int\limits_{0}^{\infty }\frac{ds}{s^{1+D/2} }\,\e^{-M^2s}\
     \prod^{D/2}_{k=1}\frac{\ff_k s }{\sinh \ff_ks}\
 \bigg[\sum_{k=1}^{D/2} \LB\cosh 2\ff_ks -1\RB -4(\prod^{D/2}_{k=1}\cosh
 \ff_ks-1) \bigg] .
\la{eexs}
\ee
This integral is UV convergent for $D\le 7$ and logarithmically
divergent for $D=8$
 implying also 
 the presence of $O(F^4)$ quadratic UV
divergence in $D=10$ SYM theory. 
It is also IR divergent for certain $\ff^{(i)}_k$  and small enough $M$
(which is a manifestation
of   the well-known tachyonic instability of the  YM theory in a
constant abelian background which is not cured  by supersymmetry).

For example,  the standard ($M=0$)
one-loop effective action for maximally  supersymmetric
$SU(2)$  YM  theory in  $D=4$  in  background $F_{mn} =  \FF_{mn} {\s_3\ov 2} $
(i.e. $\NN=1$)
is   \ci{fraa}
\begin{equation}
 \Gamma =- {4\VV_4\ov (4\pi)^2}
\int\limits_{0}^{\infty }\frac{ds}{s^3}\,
{ \ff_1 s \ov \sinh \ff_1 s}\  { \ff_2 s \ov \sinh \ff_2 s}\
(\cosh  \ff_1 s - \cosh \ff_2 s)^2  \ .
\la{yumi}
\ee

For comparison with  the matrix model expressions,
it is useful to
separate a factor  $\N \sim \sqrt {\det\ \FF_{mn}}$  in  $\Gamma_M$
representing  it as
\be
 \Gamma_M   =  \NN\  \N\  \W    \ , \ \ \la{sepa} \ee
\be
 \N \equiv
{    (2\pi)^{-D/2}} \VV_D \prod_{k=1}^{D/2} \ff_k  =
{ (2\pi)^{-D/2}} \VV_D   \sqrt {\det\ \FF_{mn}} \ ,
\la{zeee}
\ee
\be
\W = -  2
\int\limits_{0}^{\infty }\frac{ds}{s }\,\e^{-M^2 s}\
 \prod^{D/2}_{k=1}\frac{ 1 }{2\sinh \ff_ks}\
 \bigg[\sum_{k=1}^{D/2} \LB\cosh 2\ff_ks -1\RB -4(\prod^{D/2}_{k=1}\cosh
 \ff_ks-1) \bigg]\ .
\la{exs}
\ee
In the matrix model context
 $\N\inv $    will be an integer (or a rational number, cf. \rf{nuum},\rf{nuu})
 and  will be   cancelled  against  a factor  contained in $\NN$
   (see section 4).

Special cases of $\G_M$  \rf{sepa} or (up to an overall  coefficient)    $\W$
   appeared
in the discussions of interaction potentials between  D-branes (see,  e.g.,
 \ci{dkps,ikkt,ab,lifmat,lif3}).\foot{For example,
 for $\ff_1=iv,\  \ff_2,...,\ff_l=0$, \ $M=b$   we get from
\rf{exs} the
(light open string mode part of)
phase shift for the scattering of two 0-branes \ci{dkps},
$\delta=-i\W=
\int\limits_{0}^{\infty }\frac{ds}{s }\,\e^{-b^2 s} (\sin vs)\inv
 (\cos 2vs - 4 \cos vs + 3)$.}
The general expression   \rf{exs}
was given  in    \ci{CMZ}, where  it was
 describing  the potential between parallel `Dp-brane' and anti-`Dp-brane'.

The leading terms in the large $M$ expansions of $\G_M$ and $\W$
can be found directly from \rf{eexs},\rf{exs}
 $$
 \Gamma_M   = - { (3-\ha D)!\  \NN \    V_D\
\ov  (4\pi)^{D/2} M^{8-D} }\   \bigg[2\sum_{k=1}^{D/2 } \ff_k^4-(\sum_{k=1}^{D/2 }\ff_k ^2)^2\bigg]
         + O( {1 \ov M^{10-D}} )
$$
\be
  = - { (3-\ha D)! \  \NN \ V_D  \ov  (4\pi)^{D/2} M^{8-D} }
\  \bigg[ \FF_{mk} \FF_{nk} \FF_{mr} \FF_{nr}
- \fourth
(\FF_{mk} \FF_{mk})^2\bigg]    + O( {1 \ov M^{10-D}} ) \ ,
\la{tiet}
\ee
 \be
 \W   = - { (3-\ha D)! \ov 2^{D/2}  M^{8-D} }\
  { 1\ov  \sqrt {\det\ \FF_{mn}}}
\  \bigg[ \FF_{mk} \FF_{nk} \FF_{mr} \FF_{nr}
- \fourth
(\FF_{mk} \FF_{mk})^2\bigg]   + O( {1 \ov M^{10-D}} ) \ .
\la{tioet}
\ee
They  have the expected $F^4$ structure  \rf{tet}. In fact,
for the  abelian background  considered above  one has from \rf{bbb}:
\be
{\bf b}_8 = \Tr[ F^4 - \fourth (F^2)^2] =
  2\NN\ [ \FF^4 - \fourth (\FF^2)^2]
= 2\NN\ \bigg[2\sum_{k=1}^{D/2 } \ff_k^4-(\sum_{k=1}^{D/2 }\ff_k ^2)^2\bigg]\ .
\la{abel}
\ee

\newsection{ Matrix model (super Yang-Mills)  description}
In this section  we shall  demonstrate that
the leading-order terms
in the long-distance potentials  between BPS bound states
with 1/2 and 1/4  of supersymmetry
\rf{resu} and  \rf{eex}
computed  in
section 2   using classical closed string effective field theory
methods
are indeed reproduced by the instanton matrix model,
i.e.  by the  corresponding 1-loop
 SYM  computations.

The  instanton matrix model is defined by  the $D=10$  $U(N)$
SYM  Lagrangian reduced
to 0+0 dimensions (in this section we shall assume that  $T\inv = 2\pi \a'=1$)
\be
L=-\frac{1}{2g_s}{\rm tr}  \bigg(
\frac{1}{2}\lsb X_\m,X_\n\rsb^2+2\te^{T}\gamma_\m \lsb \te ,X_\m\rsb \bigg) \, ,
\label{lagran}
\ee
where  the elements of  $N\times N$ matrix $\te$
 are 16-component real spinors and $\gamma_{10}\equiv
\bI_{16\times 16}$.

We shall consider the  background gauge field
$
\bar{{ A}}_{\mu}=T\left( \bar{X}_1,\ldots
,\bar{X}_{10}\right) $
where  the components
\be
\bar{X}_i=\LB
\begin{array}{cc}
 \bar{X}_i^{(1)} & 0 \\
 0 & \bar{X}_i^{(2)} \\
 \end{array}
 \RB ,~~~~~~i=1,\ldots,8,10 \ ,
\ee
 correspond to the coordinates of the two
 BPS objects and
\be
\bar{X}_{9}=\LB
\begin{array}{cc}
 b & 0 \\
 0 & 0 \\
 \end{array} \RB
\ee
represents  their separation $b$.
The   calculation
of the  SYM one-loop  effective
action in this background  is  similar to the one described
 in \ci{CT}.
  Let us define the
 operators
\be H = \LB \xo_i \otimes { \bI}-{ \bI}\otimes
{\bar X}^{(2)*}_i\RB^2 \ , \ \ \ \ \
 H_{ij}=\xijo \otimes { \bI}+{ \bI}\otimes
{\bar X}_{ij}^{(2)*} \ ,
\ee
where  $\xijo = [\xo_i,\xo_j], \  \xijt = [\xt_i,\xt_j]$ and
  $*$ is the complex conjugation.
 The  1-loop effective action  is the sum of the
 bosonic, ghost and fermionic contributions,
\be W= W_B + W_G + W_F  \,,
\label{effact}
\ee
$$ W_B=   \ln \hbox{det} ( H
\delta_{\mu\nu}+2H_{\mu\nu}) \,, \ \ \ \ \
W_G= - 2\ln \hbox{det}\  H  \, , \ \ \ \
W_F= -\ha \ln  \hbox{det} \big(
H+\sum_{i<j}\gamma_i\gamma_jH_{ij} \big)   \ ,
$$
where the operators  act in  the $U(N)$   matrix index space, Lorentz vector
space  and Lorentz
spinor space.
In the case of the
background
$$
{\bar X}_{10} = \LB
\begin{array}{cc}
 i\de_{\tau} & 0 \\
 0 & 0 \\
 \end{array} \RB \,,  \ \ \ \ \
{\bar X}_{{ i_*}} = \LB
\begin{array}{cc}
 v\tau & 0 \\
 0 & 0 \\
 \end{array} \RB   \,,
$$
the resulting   expression for $W$ in \rf{effact}
 becomes the same   as
 found in the 0-brane matrix model \cite{CT} for
the relative motion of two BPS objects along the direction ${i_*}$.
This  may be viewed as  a manifestation of
T-duality in string theory  or Eguchi-Kawai reduction in
large $N$  SYM theory \ci{tay,grt,ikkt}.

\subsection{D-instanton -- `Dp-brane' interaction
}
A  `Dp-brane'  wrapped over a
 torus $\td T^{p+1}$
is represented by the following classical solution  of  the
 instanton matrix model ($m,n= 1,...,p+1=2l$)
\be
\bar X_m = T\inv (i  \del_m + \td A_m)
{\bI}_{n_{-1}\times n_{-1}}
\ , \ \ \ \ \  \ \
[\bar X_m, \bar X_n] = i T^{-2}
\td F_{mn}{\bI}_{n_{-1}\times
 n_{-1}} \ , \la{som}
\ee
where $  \del_m$ act on functions on
the  torus and  $ \td F_{mn}$  is  a constant abelian field
strength.  This configuration corresponds  \ci{dbi} to the
$(p+(p-2)+\cdots +1+i) $  type IIB bound state
wrapped over the torus $ T^{p+1}$ dual to $\td T^{p+1}$.
We shall choose $\td F_{mn}$ in the form
\begin{equation}
  \td \F_{mn}\equiv T\inv \td F_{mn}=\left(
 \begin{array}{ccccc}
 0 & \f_1 & & & \\
 -\f_1 &0& & & \\
  & &\ddots & & \\
 &&&0 & \f_l \\
 &&&-\f_l&0 \\
 \end{array}
 \right).
 \label{cmunu}
 \end{equation}
This background field is (minus) the inverse of the one
which appears in  the  T-dual   string theory picture \rf{fmunu},
i.e. $\FI_{mn} \F_{m'n} = \d_{mm'}$, or $\  \f_k =  f_k\inv$.

Let us explain
  the reason for this inverse identification
 between the fluxes in the  matrix model  and  string theory
descriptions (see \ci{grt,lifmat} for discussions of  T-dual type IIA  cases).
The   $U(N)$  SYM theory
on  $ T^{p+1}$represents $n_p=N$ \   Dp-branes  with  euclidean world-volumes
 wrapped over the torus   \cite{wit2}.
By T-duality \ci{tay}, it is also describing
$\td n_{-1}=N$ \  D-instantons on the dual torus $\td T^{p+1}$.
Turning on the background field \rf{fmunu} on
$ T^{p+1}$    we get
a non-marginal bound state $p+ (p-2)+ ...+i$
with the  `induced' instanton number  \rf{number},\rf{nuum}
  equal to
  $ n_{-1} =n_p V_{p+1} ( {Tf\ov 2\pi})^{p+1\ov 2}$
(for simplicity  here we  set
all $f_k$ to be equal to $f$).
Since T-duality along all of the directions of the torus  $ T^{p+1}$
interchanges instantons with Dp-branes,
the corresponding  bound state  wrapped over
 $\td T^{p+1}$  contains
$\td n_p =  n_{-1}$  Dp-branes and $\td n_{-1} =n_p$
instantons.
If the background field $\td \F_{mn}$
that produces this charge distribution
is   \rf{cmunu}, then
$\td  n_{-1} =\td n_p \td V_{p+1} ({T\td f\ov 2\pi})^{p+1\ov 2} .$
As a result,
\be
\td n_p =  n_{-1} \ , \ \ \ \   \  \td n_{-1} =n_p \
\la{eqaa}
\ee
 implies
\be
 V_{p+1}  \big({Tf\ov 2\pi}\big)^{p+1\ov 2} \
 \td V_{p+1}\big({T\td f\ov 2\pi}\big)^{p+1\ov 2} =1 \
\ , \ \ \ \ {\rm i.e.} \  \  \ \ \ \ \  f\  \td f   =  1
\ , \la{vvv}
\ee
where we used \rf{voo}.

The  matrix model background describing  the
 configuration of `Dp-brane' ($p=2l-1$)
 with the world-volume directions
$X_1,...,X_{p+1}$   and $N_{-1}$ D-instantons  located at
the origin,
  which are
separated  from each other
 by a distance $b$ in the $9$-th direction is thus
represented by
\be
\xo_1= q_1 \,,  \ \
\xo_2= p_1 \,, \ ... \ , \ \ \ \
\xo_{2l-1}= q_{l} \,, \  \
\xo_{2l}= p_{l} \,, \ \ \ \ \  \xo_9= b \,, \label{GFBK}
\ee
\be
[q_k,p_n]=i\f_k \delta_{kn}{ \bI }   \ , \ \ \ \ \ \ \ \f_k= f\inv_k
 \ ,
\la{rele}
\ee
with all other $\xo_\m$ components being  equal to zero and
the ${N_{-1}\times N_{-1}}$ matrix
$\xt_{\mu}$ ($\m=1,...,10$)  having zero entries.

The bosonic, ghost and fermionic contributions to the 1-loop
effective action $W$
\rf{effact} in this background   are
\be
W_B= N_{-1}\sum_{\{ n\} =0}^{\infty}
\Tr\ \hbox{ln}\LB  { b}_{\{ n\}}^2
\delta_{\mu\nu}   - 2i\FI_{\m\n} \RB \,,
\ \ \ \ \ W_G=-2N_{-1}\sum_{\{ n\} =0}^{\infty}
\Tr\ \hbox{ln}\  { b}_{\{n\} }^2   \,,
\ee
\be
W_F=-\ha {N_{-1}}\sum_{\{ n\} =0}^{\infty}
\Tr\ \hbox{ln} \bigg( { b}_{\{ n\}}^2
+\frac{i}{2}\gamma_{ij}\FI_{ij} \bigg) \,,
\ \ \ \ \ \ { b}_{\{ n\}}^2\equiv b^2+\sum^l_{k=1}  \f_k (2n_k+1) \,,
\ee
where  $\{ n\}\equiv \{ n_1,...,n_l\}$.
The constant  background field matrix
 $\FI_{\m\n}$  has  \rf{cmunu}
 as non-zero entries.
 The  resulting  effective action $W$ \rf{effact}
is given by
\begin{equation}\label{otvet}
 W=-2N_{-1}\int\limits_{0}^{\infty }\frac{ds}{s}\,\e^{-b^2s}
 \prod^l_{k=1}\frac{1}{2\sinh \f_ks} \bigg[\sum_{k=1}^l \LB\cosh 2\f_ks -1\RB -4(\prod^l_{k=1}\cosh
 \f_ks-1) \bigg]\ .
\la{pio}
 \end{equation}
This   is { equal} to
the   1-loop  effective action $\G_M$
of the $U(N)$    SYM theory on the dual torus $\td T^{p+1}$
 in a constant
abelian   background proportional to
$\FI_{mn}$  and   with   an
IR cutoff $M=b$ (see \rf{gam},\rf{eexs},\rf{sepa}).
The  dimension of the fundamental
representation of the YM gauge group
is the total number of instantons  $N= N_{-1} + n_{-1}$
(with both  $N_{-1}$ and $n_{-1}$  assumed to be large).

Indeed,    let us  set
$ \  D=p+1=2l, \ \ff_k =  \f_k, \  \VV_D= \td V_{2l}$
and
\be
N= N_{-1} + n_{-1} \ , \ \ \ \ \  \ \
\NN=  {n_{-1}} N_{-1} \la{nnnn}
\ee
in the
  SYM  expression \rf{eexs} or \rf{sepa},\rf{exs}.
The corresponding abelian  $U(N)$ SYM background  in the fundamental representation is
given
by a diagonal $N \times N$ matrix
$
 F_{mn} =\LB
\begin{array}{cc}
\FI_{mn} \ {\bI}   & 0 \\
 0 &  0  \\
 \end{array} \RB$
where ${\bI}$ is a unit ${n_{-1}} \times {n_{-1}}$ matrix.
 In the adjoint
representantion it has  $\NN=n_{-1} N_{-1}$  non-zero entries
 $(\FI_{mn},\ -\FI_{mn} )$  (differences of diagonal
values  of the Cartan subalgebra element
in the fundamental representation).
Equivalently,
$\NN= q({N_{-1} + n_{-1}})  - q({N_{-1}})  - q({ n_{-1}})
 =  {n_{-1}} N_{-1}$, where $q(N) = \ha N(N-1)$
is the number of positive roots of $U(N)$.
The resulting SYM effective action is thus given by \rf{eexs}.

Since the factor $\N$ in \rf{zeee}  on  the dual torus  is
$\N= {\td n_{-1} \ov \td n_p}= { n_{p} \ov  n_{-1}}$,
we conclude that  for $n_p=1$
(assumed in the derivation of   \rf{pio})  the factor of   $\N={ 1\ov n_{-1}} $
cancels out, i.e.
\be
 \G_M  =\N\  \NN\  \W=   N_{-1}\ \W = W \ . \ee

Retaining only the leading term in the large distance
($b\to \infty$) expansion of $W$, we  find  the same expression as
 in  \rf{tiet},\rf{tioet}
 \be
 W=-\,
\frac{1}{b^{8-2l}}2^{-l} (3-l)!\, N_{-1} \prod_{k=1}^{l }\f_k ^{-1}
\
 \bigg[2\sum_{k=1}^{l } \f_k^4-(\sum_{k=1}^{l }\f_k ^2)^2\bigg]
 +O (\frac{1}{b^{10-2l}}) \,.
\la{pott}
 \ee
Remarkably, with  $b=r$ and $ \f_k= f\inv_k$ this
 coincides
with   the long-distance interaction potential \rf{resu},\rf{rsu},\rf{rsut}
found from supergravity
 in the limit of   {\it large}
 instanton  number $n_{-1}$  (large  $f_k$ or small  $\f_k$).

The coefficient of the subleading $\frac{1}{b^{10-2l}}$ term in \rf{pott}
turns out  to be  zero.
This is   a consequence of the vanishing of the
coefficient   ${\bf b}_{10}$  \rf{bib} in  \rf{poop}
in a constant abelian background.
Note, however,  that the powers of $r=b$ in the  subleading
terms in \rf{pio},\rf{pott}  and in the supergravity expression  \rf{saa}
do not match  in general.

 The same universal expressions \rf{pott} or \rf{abel}
 describe also  interactions of T-dual configurations of
branes in  the  0-brane matrix model.
For example, the scattering of the two 0-branes  is represented
 by  the  (electric)
background $F_{01} = iv$,  $N=n_0 + N_0, \ \NN= n_0 N_0$, i.e.
$l=1, \ \f_1=iv, \ \delta = -i W  \sim  {1 \over  r^6} v^3$.
The case  of a 0-brane scattering on a $2+0$ brane
 is represented by  $N\times N$
 matrix
$
 F_{mn} =\LB
\begin{array}{cc}
\FI^{(1)}_{mn} \ {\bI}_{n_0 \times n_0}   & 0 \\
 0 &  \FI^{(2)}_{mn} \ {\bI}_{N_0 \times N_0} \\
 \end{array} \RB,$
where   $\FI^{(1)}_{mn} = \f \epsilon_{mn}$ for $ m,n=2,3$, \ 
$\FI^{(2)}_{mn} = iv \epsilon_{mn}$ for $ m,n=0,1$
($1$ is  dual to the direction of the 0-brane motion,
$2,3$ are dual to the directions of 2-brane) and we assume that 
the volume $\td V_2$ of the two-torus $(\td x_0,\td x_1)$ 
 is chosen so that  $iv \td V_2 =2\pi$.
  In the adjoint representation  this background is given by
$ F_{mn} =
   {\rm diag} (0, ...,0,\FF_{mn} , - \FF_{mn}, ..., \FF_{mn}, - \FF_{mn}) $
with the total number of non-zero entries $2\NN =2n_0 N_0$
and $\FF_{mn} = \FI^{(1)}_{mn} - \FI^{(2)}_{mn}$ (which has block-diagonal structure as the non-zero components of $\FI^{(1)}_{mn}$ and $\FI^{(2)}_{mn}$
are orthogonal).
As a result,  here  $l=2, \ \f_1= \f, \ \f_2=iv$
and thus   $\delta = -i W  \sim
{ 1\over v \f r^4} ( \f^4 + 2 v^2 \f^2 + v^4)$
in agreement with \ci{lifmat}.

\subsection{Interaction of  `D-string'   with  3-brane--instanton bound state
}
The matrix model  background corresponding to   the configuration
 of the  `D-string'  ($1+i$)
wrapped over a 2-torus in  $(5,6)$ directions  and the D3-brane--D-instanton
bound state $(3\pa i)$ wrapped over a 4-torus in $(1,2,3,4)$
directions,  which are  separated by a  distance $b$
 in the $9$-direction is given by ($a=1,...,4$; \ $T=1$)
\be
\xo_a= T\inv(   i \de_a+ \td A_a)  =P_a \,,   \ \ \ \  \ \
\xo_9= b \,,
\label{MFBK}
\ee
$$
\xt_5 =q \,, \ \ \ \
\xt_6 = p \, , \ \ \ \ \ \ \  [q,p] =i \f { \bI}  \ ,  $$
where  the   $U(N_{-1})$ gauge potential  $\td A_a$ is  representing  the charge
$N_3$    instanton on the dual torus $\td T^4$ (see, e.g.,  \ci{grt}),
 i.e. its
 field strength
$
G_{ab} = \del_a \td  A_b - \del_b \td A_a -i [  \td A_a, \td A_b] $
satisfies
\be
G_{ab} =*G_{ab} \ , \ \ \ \ \ \ \ \ \
{1\ov 16\pi^2   } \int_{\td T^4} d^4x {\ \rm tr}( G_{ab}G_{ab}) =  N_3 \,.
\la{value}
\ee
The bosonic, ghost and fermionic  contributions to the effective action
\rf{effact} in this background   are
\be
W_B= \sum_{n=0}^{\infty}
\Tr\ \hbox{ln}\lsb ( { b}_n^2+P^2 )
\delta_{\mu\nu} - 2i \F_{\m\n}^{\prime} \rsb \,,
\ \ \ \ \  \
W_G=-2\sum_{n=0}^{\infty}
\Tr\ \hbox{ln} \LB { b}_n^2 + P^2 \RB  \,,
\ee
$$
W_F=-\ha \sum_{n=0}^{\infty}
\Tr\ \hbox{ln} \bigg( { b}_n^2
+P^2+\frac{i}{2}\gamma_{mk} \F_{mk}^{\prime}  \bigg) \,, \ \ \ \
\ \ {b}_n^2\equiv b^2+\f (2n+1)\  .
 $$
where
\be
 \F_{mn}^{\prime} =\LB
 \begin{array}{cc} G_{ab} & 0 \\ 0 &
\FI_{\a\b}  \\
 \end{array} \RB  ,  \ \ \ \ \ \ \
 \FI_{\a\b}=\f\ep_{\a\b}\ .
\ee

The expression for $W$ is computed in a similar way as in
the T-dual  case   of $(2+0)$--$(4\pa 0)$ configuration
considered in  \ci{CT}.
The final result for the leading long-distance $(b\to \infty$)
term in  $W$ is
\be
W= \frac{1}{32\pi^2b^2}
\bigg[ \f \int_{\td T^4}  d^4x {\ \rm tr} (G_{ab} G_{ab})
-\td V_4N_{-1}\f^3 \bigg] + O( { 1 \ov b^4})
 \la{eexi}
\ee
$$
 = \frac{1}{2b^2} \bigg(N_3 \f - {\textstyle{ {1\ov 16 \pi^2}}} \td V_4N_{-1} \f^3\bigg)  + O( { 1 \ov b^4})
   \ .
$$
This  becomes exactly  the same as   the supergravity result
for the interaction potential \rf{eex}
after  we set $b= r, \ \f= f\inv, \ n_1=1$, use the relation \rf{voo}, i.e.
$\td V_4 V_4 = (2\pi)^4$,
and note  that since it is assumed that
  $ N_{-1} \gg N_3$ the last term in \rf{eex}
can be neglected.

The expression \rf{eexi}  is {equivalent}  to the
leading-order  $O(F^4)$ term in the $U(n_{-1} + N_{-1})$
  SYM effective action \rf{tet}  on
 the dual 6-torus $ \td T^2 \times \td T^4$
computed
in the  background
 \be
 F_{mn} = \hat \F_{mn} =\LB
 \begin{array}{cc} G_{ab} & 0 \\ 0 &
\FI_{\a\b}{\bI}_{n_{-1}\times n_{-1}}  \\
 \end{array} \RB\,. 
\label{bacc}
\ee

Indeed,
substituting  $\hat \F_{mn}$  \rf{bacc}
into \rf{fund}, i.e.  into   ${\bf b}_8$
in \rf{bbb},
and observing  that the $G^4$-terms cancel out (${\bf b}_8$
 vanishes
on  a  self-dual  gauge field background)
one is  left with  the  the abelian $\FI^4$ term and the  `cross-term'
$\FI^2 G^2$,  i.e.
\be
{\bf b}_8 (\hat \F)  =
{\rm Tr} \bigg[ \FI^4 - \four (\FI^2)^2  -  \ha \FI^2 G^2 \bigg] =
2n_{-1} \bigg[  N_{-1}  \f^4 -  \f^2 {\rm tr} (G_{ab} G_{ab})\bigg]   \ ,
\la{ree}
\ee
where in the first expression $ \FI$ and $G$
are the adjoint representation counterparts
 of the   $N\times N$   matrices  in the fundamental representation
with  non-vanishing
$n_{-1} \times n_{-1}$   and  $N_{-1} \times N_{-1}$
blocks (note that the spatial components of $ \FI$ and $G$
are orthogonal).
One can also derive this expression by formally  representing  $G_{ab}$
 as an abelian matrix   with   two equal $2\times 2$ blocks, i.e.
$G_{ab} = g \epsilon_{ab}
 {\bI}_{N_{-1}\times N_{-1}}$ for $a,b=1,2$ and $a,b=3,4$.
Then  one may apply  formula  \rf{abel} for the abelian  background
with $\ff_1 = \f,\  \ff_2 = \ff_3=g$. This gives
$ {\bf b}_8 = 2 n_{-1} N_{-1}
 \big[ 2(\f^4 + 2 g^4)
-(\f^2  + 2 g^2)^2\big] = 2 n_{-1} N_{-1}
 \big( \f^4  - 4 \f^2 g^2 \big)$, i.e.
 the same result as in \rf{ree} since
${\rm tr} (G_{ab} G_{ab}) = 4 N_{-1} g^2$, \ $(2\pi)^{-2} \td V_4 g^2 N_{-1} = N_3$ (cf. \rf{number}).

For $n_1=1$ one has
$n_{-1} =  2\pi \td V_2\inv \f\inv  $  (see \rf{nuu})
 and  concludes that
\be
\G_M = - { 1 \ov 2 (4\pi)^3 M^2  } \td V_2 \int_{\td T^4}  d^4x\   {\bf b}_8 + O( {1 \ov M^4})
\ee
 in \rf{tet}
is  equal to  $W$ in \rf{eexi} for $b=M$.
This is also  in agreement  with
  the supergravity potential represented in the form \rf{erx}.


Thus we have found  complete agreement between the  1-loop
matrix model and classical supergravity
expressions  for the leading-order long-distance interaction potentials.

\medskip
\centerline {\ \bf Acknowledgments}
We are grateful to Yu. Makeenko for  useful discussions.
The work of I.C. was supported in part by CRDF grant 96-RP1-253.
A.A.T.  acknowledges  the support
 of PPARC and  the European
Commission TMR programme grant ERBFMRX-CT96-0045.

{\vspace*{\fill}\pagebreak}


\begin{thebibliography}{99}
\addtolength{\itemsep}{-6pt}

\bibitem{ikkt}
N.~Ishibashi, H.~Kawai, Y.~Kitazawa and A.~Tsuchiya,
{\it A Large--N Reduced Model as Superstring}, hep-th/9612115.

\bi{peri}
V. Periwal, {\it Matrices on a point as the theory of everything},
\pr D55 (1997) 1711,
 hep-th/9611103.

\bi{dbi} A.A. Tseytlin, {\it On non-abelian generalisation of
Born-Infeld action in string theory}, hep-th/9701125.

\bi{li} M. Li, {\it Strings from IIB matrices}, hep-th/9612222.

\bibitem{bfss}
 T. Banks, W. Fischler, S.H. Shenker and L. Susskind,
\pr D55 (1997) 5112, hep-th/9610043.

\bibitem{CMZ}
I.~Chepelev, Y.~Makeenko and  K.~Zarembo,
{\it Properties of D-Branes in Matrix Model of IIB Superstring},
hep-th/9701151.

\bi{fas}
A.  Fayyazuddin and D.J. Smith,
{\it  p-brane solutions in IKKT IIB matrix theory},
hep-th/9701168.

\bi{FMOSZ}
A. Fayyazuddin, Y. Makeenko, P. Olesen, D.J. Smith and K. Zarembo,
{\it Towards a non-perturbative formulation of IIB superstrings by matrix models}, hep-th/9703038.



\bibitem{grt}
O.J.~Ganor, S.~Ramgoolam and W.~Taylor,
{\it Branes, Fluxes and
Duality in M(atrix) Theory}, hep-th/9611202.

\bibitem{bss}
T.~Banks, N.~Seiberg and S.~Shenker,
{\it Branes from Matrices}, hep-th/9612157.

\bibitem{lifmat}
G. Lifschytz and S.D. Mathur,
{\it Supersymmetry and Membrane Interactions in M(atrix) Theory},
hep-th/9612087.
\bibitem{lif3}
G. Lifschytz,
{\it Four-Brane and Six-Brane Interactions in M(atrix) Theory},
hep-th/9612223.


 \bibitem{wit2}
E. Witten, {Nucl. Phys.} {B460} (1995) 335, hep-th/9510135.

\bi{ggu}
M.B. Green and M. Gutperle,
\pl B398 (1997) 69,
  hep-th/9612127.

\bi{ggut}
M.B. Green and M. Gutperle, Nucl. Phys. B476 (1996) 484, hep-th/9604091.

\bi{tsee}
A.A. Tseytlin, \prl 78 (1997) 1864, hep-th/9612164.



\bibitem{ab}
O. Aharony and M. Berkooz,
{\it Membrane Dynamics in M(atrix) Theory},
hep-th/9611215.

\bibitem{CT}
I.~Chepelev and  A.A.~Tseytlin,
{\it Long-distance  interactions of D-brane  bound states
and longitudinal 5-brane in M(atrix) theory },
hep-th/9704127.

 \bi{dps}
 M. Douglas,  J. Polchinski  and A.  Strominger,
 {\it Probing Five-Dimensional Black Holes with D-Branes},
 hep-th/9703031.


\bi{mall}
J. Maldacena, {\it Probing near extremal black holes with D-branes}, hep-th/9705053.


\bi{lim} M. Li and E. Martinec,
 {\it Matrix black holes}, hep-th/9703211;
  {\it On the entropy of matrix black holes}, hep-th/9704134.

\bi{dvv}
R. Dijkgraaf, E. Verlinde and  H. Verlinde,
{\it 5-D Black holes and matrix strings}, hep-th/9704018.

\bi{hal}
E. Halyo,  {\it M(atrix) black holes in five dimensions}, hep-th/9705107.

\bibitem{pol}
J.~Polchinski, Phys.\ Rev.\ Lett. {75} (1995) 4724;
 {\it TASI Lectures on D-Branes}, hep-th/9611050.


\bibitem{doug} M.R. Douglas, {\it
 Branes within Branes},  hep-th/9512077.

\bi{ggp}
G.W. Gibbons, M.B.  Green and M.J. Perry, \pl B370 (1996) 37, hep-th/9511080.


  \bi{ts2}
 A.A. Tseytlin, \np B475 (1996) 149,  hep-th/9604035.


\bi{fraa} E.S. Fradkin and A.A. Tseytlin, \np B227 (1983) 252.


\bi{mets} R.R. Metsaev  and A.A. Tseytlin, \np B298 (1988) 109.

\bi{vdv} A.
E.M. Van de Ven, \np B250 (1985) 593.

\bi{okubo}  S. Okubo and J. Patera, \pr D31 (1985) 2669.

\bi{tgw}
 A.A. Tseytlin, \np B276 (1986) 391;
D. Gross and E. Witten, \np B277 (1986) 1.

\bi{grs}
M.B. Green and J. Schwarz, \np B198 (1982) 441;
M.B. Green, J. Schwarz and L. Brink, \np B198 (1982) 474.


\bi{ber}
D. Berenstein and R. Corrado, {\it M(atrix) theory in various dimensions},
hep-th/9702108.

\bi{avr}
I.G. Avramidi, Journ. Math. Phys. 36 (1995) 1557, gr-qc/9403035; hep-th/9604160.

\bibitem{dkps}
M.R. Douglas, D. Kabat, P. Pouliot and S.H. Shenker,
 \np B485 (1997) 85,
hep-th/9608024.


\bibitem{tay}
W. Taylor,   \pl B394 (1997) 283,
hep-th/9611042.


\end{thebibliography}
\end{document}